\documentclass{article}
\usepackage{spconf,amsmath,graphicx}
\usepackage[numbers,sort&compress]{natbib}

\usepackage{epsfig}
\usepackage{amsfonts}
\usepackage{algorithm}
\usepackage{algorithmic}
\usepackage{multirow}
\usepackage{url}

\usepackage{hyperref}

\title{Meta-Voice: Fast few-shot style transfer for expressive voice cloning using meta learning}

\name{Songxiang Liu, Dan Su, Dong Yu}
\address{Tencent AI Lab}

\begin{document}
%
\maketitle
\begin{abstract}
The task of few-shot style transfer for voice cloning in text-to-speech (TTS) synthesis aims at transferring speaking styles of an arbitrary source speaker to a target speaker's voice using very limited amount of neutral data.
This is a very challenging task since the learning algorithm needs to deal with few-shot voice cloning and speaker-prosody disentanglement at the same time.
Accelerating the adaptation process for a new target speaker is of importance in real-world applications, but even more challenging.
In this paper, we approach to the hard fast few-shot style transfer for voice cloning task using meta learning. We investigate the model-agnostic meta-learning (MAML) algorithm and meta-transfer a pre-trained multi-speaker and multi-prosody base TTS model to be highly sensitive for adaptation with few samples. Domain adversarial training mechanism and orthogonal constraint are adopted to disentangle speaker and prosody representations for effective cross-speaker style transfer.
Experimental results show that the proposed approach is able to conduct fast voice cloning using only 5 samples (around 12 second speech data) from a target speaker, with only 100 adaptation steps. Audio samples are available online \footnote{\url{https://liusongxiang.github.io/meta-voice-demo/}}.

\end{abstract}
\section{Introduction}
\label{sec1:intro}

The quality, naturalness, and intelligibility of generated speech by end-to-end neural text-to-speech (TTS) synthesis models have been greatly improved in the past few years \cite{wang2017tacotron,shen2018natural, yu2019durian}.
Nowadays, it is in high demand to adapt an existing TTS model to support personalized and multi-prosody speech synthesis for a new target speaker using his/her limited amount of data (e.g., 5 utterances).
We refer to this task as few-shot style transfer for voice cloning in TTS, aiming at transferring speaking style of an arbitrary source speaker to a target speaker's voice under data-scarcity constraint. 
In real-world applications, it is promising to endow a TTS model with the capability of conducting fast adaptation, meaning that the users can obtain instantaneous personalized speech interface.
The task of fast few-shot style transfer is very challenging in the sense that the learning algorithm needs to deal with not only a few-shot voice cloning problem (i.e., cloning a new voice using few samples) but also a speaker-prosody disentanglement and control problem. 

\begin{figure}[t]
	\centering
	\includegraphics[width=0.5\textwidth]{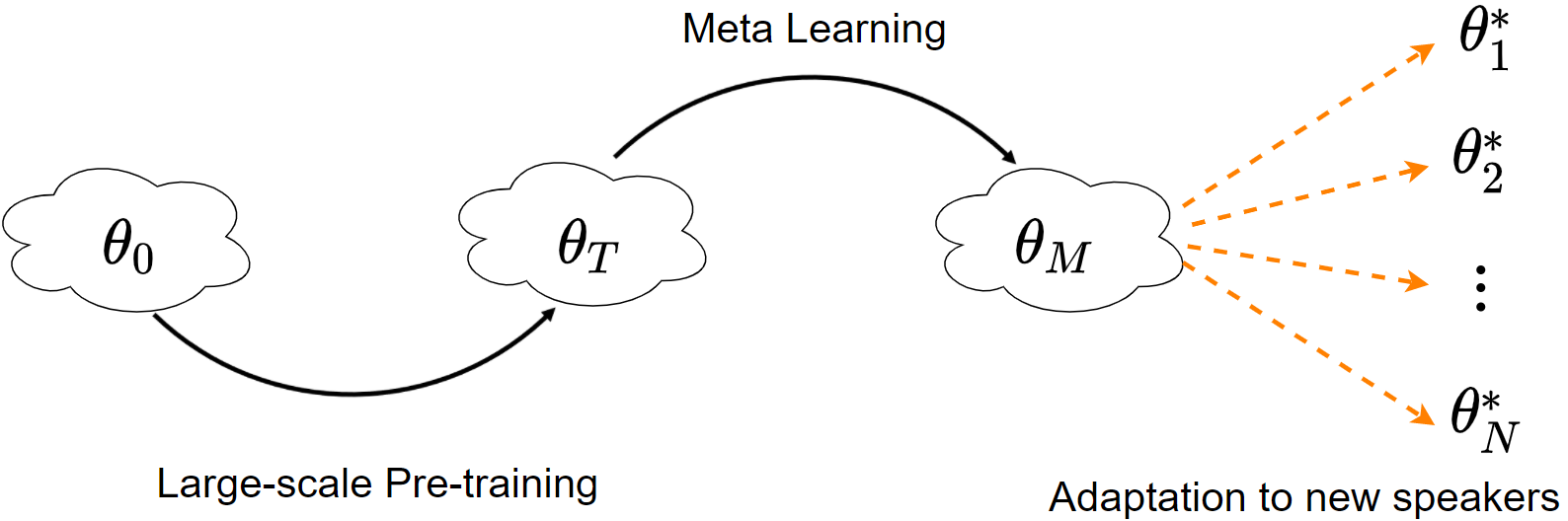}
	\caption{Illustration of training procedure of Meta-Voice, which contains three stages of large-scale pre-training, meta learning and adaptation to new speakers.}
	\label{fig:overview}
\end{figure}

Few-shot voice cloning alone is a challenging task, due to the reason that the trade-off between model capacity and overfitting issues must be well dealt with. 
Existing approaches can be roughly classified into two kinds \cite{NEURIPS2018_4559912e}, i.e., speaker encoding approach and speaker adaptation approach, in terms of whether to update parameters of the base TTS model during voice cloning.
Speaker encoding approach does not update parameters of the base TTS model and represents speaker characteristics using speaker embedding vectors, which are expected to control the output voice. This approach does not involve a fine-tuning process and conducts voice cloning in a zero-shot adaptation manner. In \cite{jia2018transfer}, a speaker encoder model is first separately trained with a speaker verification loss \cite{wan2018generalized} leveraging large-scale low-quality data; and then adopted to compute speaker embedding vectors for a multi-speaker TTS model. During voice cloning, the speaker encoder computes an embedding vector from the limited adaptation data for a new speaker and adapts the base TTS model to the new voice by using the embedding vector as conditioning. Instead of pre-training a speaker verification model using external data, the speaker encoder model can also be trained using the TTS data, either jointly optimized with the base TTS model or separately optimized \cite{NEURIPS2018_4559912e, chen2018sample, min2021meta}. This kind of approaches well tackle the possible overfitting issue since there is no finetuning process. However, embedding vectors computed by a pre-trained speaker encoder may not generalize well to an unseen speaker, which further degrades the voice cloning performance.
Speaker adaptation approach \cite{NEURIPS2018_4559912e, chen2018sample, chen2021adaspeech} finetunes all or part of the base TTS model with limited adaptation data, which rely on careful selecting weights that are suitable for adapting and adopting proper adaptation scheme to update these weights in order to avoid overfitting the adaptation data. However, the cloning speed is much slow than the speaker encoding approaches \cite{chen2021adaspeech}.

In this paper, we aim at tackling the more challenging fast few-shot style transfer problem, where we not only make few-shot voice cloning faster but also enable the TTS model to transfer from source styles to the cloned voice.
To make the voice cloning procedure faster, we extend the optimizing-base model-agnostic meta-learning (MAML) approach \cite{finn2017model} to TTS.

To avoid overffiting issues when updating all model parameters with limited adaptation data, we split model parameters into style specific parameters and shared parameters. In what follows, we term the proposed model as Meta-Voice for clarity. 
As illustrated in Fig.~\ref{fig:overview}, we transfer the shared parameters from a multi-speaker and multi-prosody base TTS model pre-trained with a large-scale corpus and conduct meta-training on the style specific parameters to a state which is highly sensitive for fast adaptation with only few samples from a new speaker.
For the purpose of style transfer, we adopt domain adversarial training mechanism \cite{ganin2016domain} and orthogonal constraint \cite{bian2019multi} to disentangle speaker and prosody representations.
The contributions of this work are summarized as follows:
\begin{itemize}
    \item To the best of our knowledge, this paper is the first work addressing the challenging few-shot style transfer problem for voice cloning.
    \item We use meta-learning algorithms on a pretrained TTS model to make it adaptable to a new voice quickly with few samples.
    \item We adopt effective speaker-prosody disentangling methods to achieve few-shot cross-speaker style transfer.
\end{itemize}

The rest of this paper is organized as follows:
Details of the proposed approach are presented in Section~\ref{sec2} and Section~\ref{sec3}. Experimental results are shown in Section~\ref{sec4} and Section~\ref{sec5} concludes this paper.

\begin{figure}[t]
	\centering
	\includegraphics[width=0.45\textwidth]{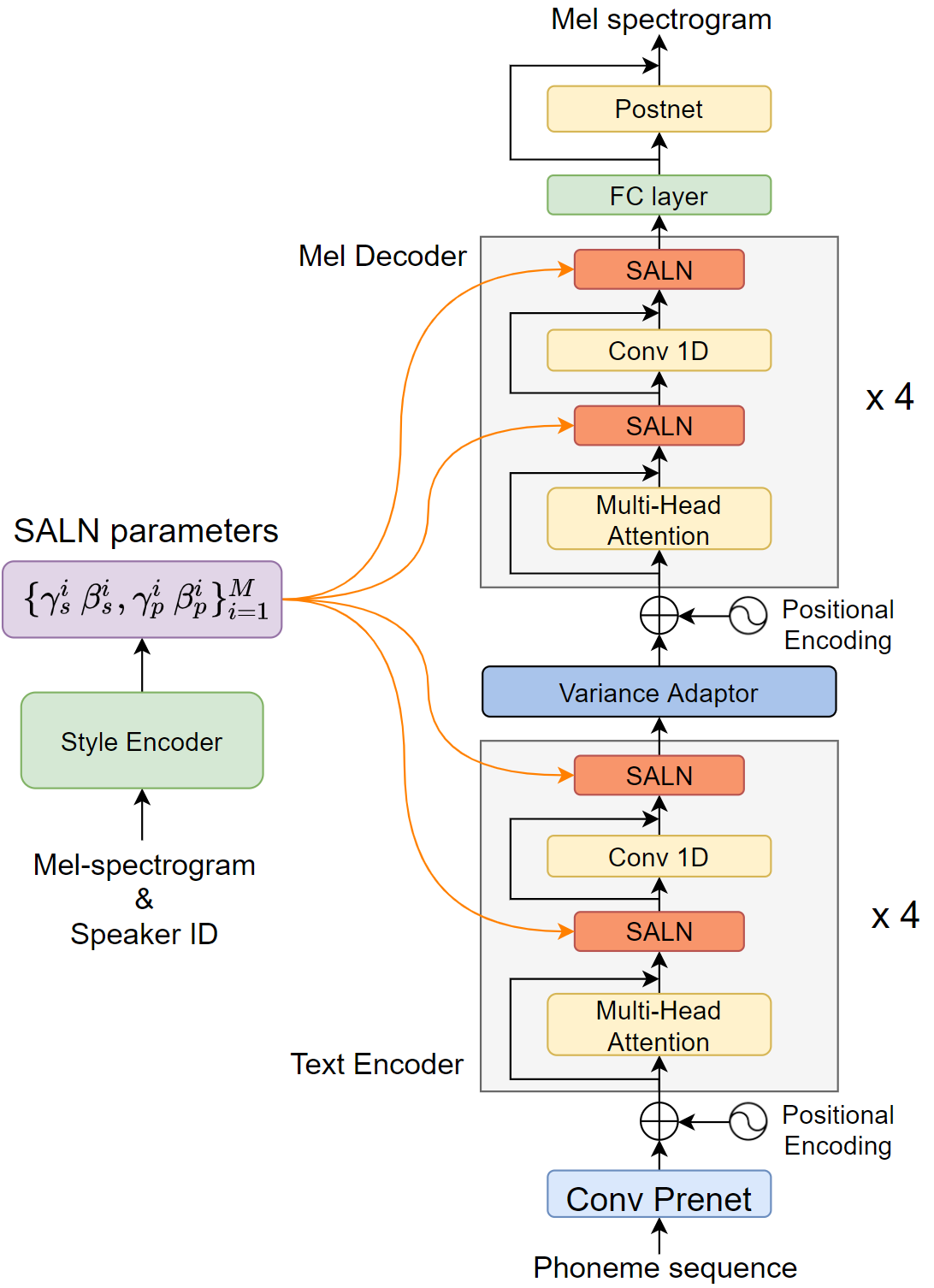}
	\caption{The base model architecture. SALN represents ``style-adaptive layer normalization", whose details are illustrated in Fig.~\ref{fig:saln}. In total, $M$ SALN layers are used in the base model. Details of the style encoder are illustrated in Fig.~\ref{fig:style_encoder}.}
	\label{fig:model}
\end{figure}

\begin{figure}[t]
	\centering
	\includegraphics[width=0.45\textwidth]{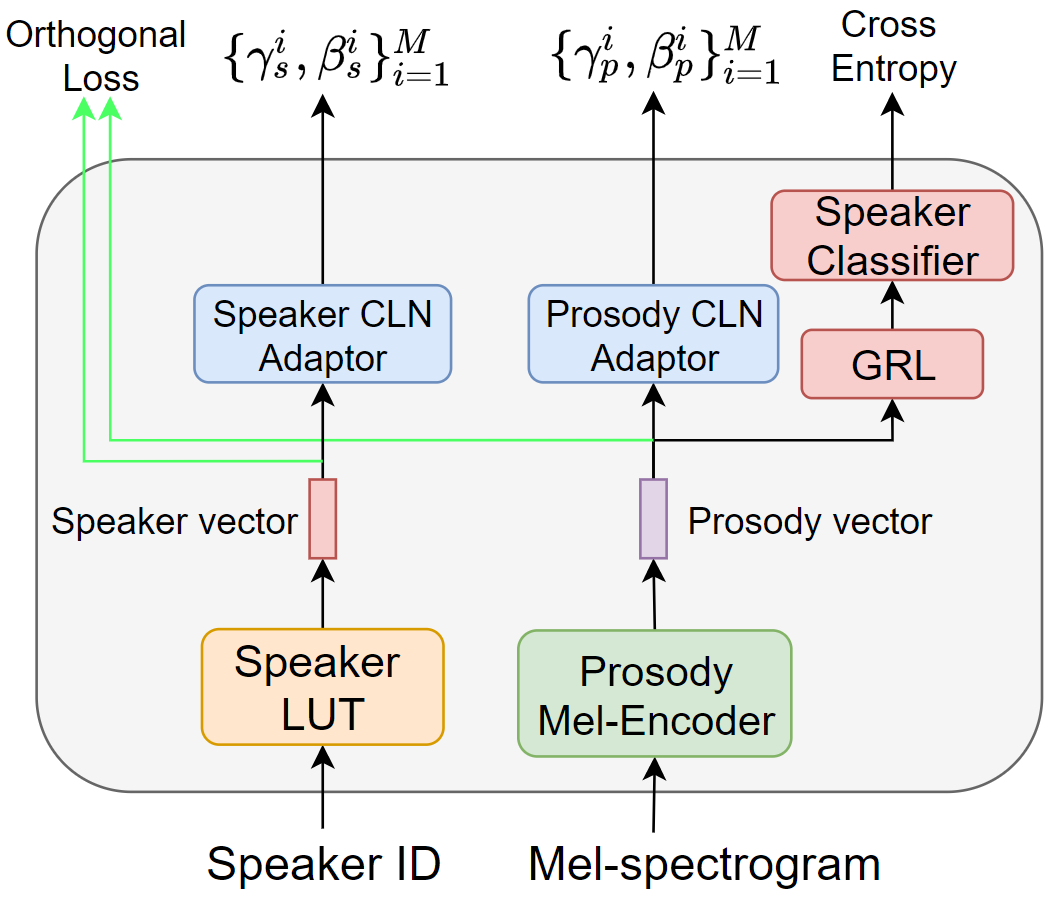}
	\caption{Style Encoder. GRL represents ``gradient reversal layer" and LUT is for ``look-up table". $M$ is the total number of SALN layers in Meta-Voice.}
	\label{fig:style_encoder}
\end{figure}

\begin{figure}[t]
	\centering
	\includegraphics[width=0.38\textwidth]{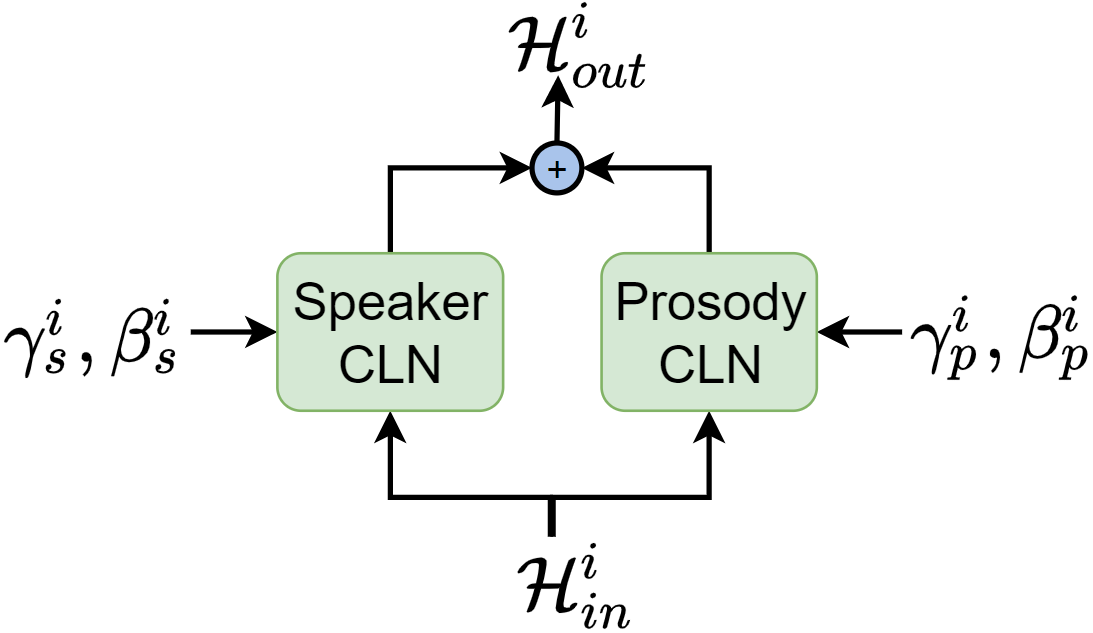}
	\caption{The style-adaptive layer normalization. CLN represents ``conditional layer normalization". $\mathcal{H}_{in}^i$ is the input hidden feature of the $i$-th SALN layer, whose output hidden feature is $\mathcal{H}_{out}^i$.}
	\label{fig:saln}
\end{figure}


\section{Model architecture of Meta-Voice}
\label{sec2}

In this section, we describe the architecture of Meta-Voice for multi-speaker and multi-style speech generation.
As shown in Fig.~\ref{fig:model}, Meta-Voice consists of four major modules: a text encoder, a variance adaptor, a mel decoder and a style encoder.
In this work, we decouple a speaking style into speaker and prosody components, omitting other factors such as phonetics and channel effects, according to the ``subtractive" definition of prosody in \cite{skerry2018towards}.
We extend the style-adaptive layer normalization (SALN) presented in \cite{min2021meta} to support multiple speakers and prosodic types and add SALN layers in both the text encoder and the mel decoder, making Meta-Voice fully conditioned on speaking style information.
We split parameters $\theta$ in Meta-Voice into shared parameters $\bar{\theta}$, speaker-related parameters $\theta^{s}$ and prosody-related parameters $\theta^{p}$, i.e., $\theta=\{\bar{\theta}, \theta^s, \theta^p\}$.
$\bar{\theta}$ contains parameters of the text encoder, the variance adaptor and the mel decoder, which are responsible of producing right pronunciations corresponding to the input textual content. $\theta^{s}$ and $\theta^{p}$ as a whole contain parameters of the style encoder, which generates scaling and shifting parameters for SALN layers, modulating hidden features in both the text encoder and mel decoder to control timbre and prosodic characteristics of generated speech.

\subsection{Shared model parameters}

The text encoder is the same as the one presented in \cite{min2021meta}, except that we use different SALN layers. We convert normalized text into phoneme sequence, which are transformed into trainable real-valued text embeddings with a text embedding layer.
A convolutional prenet, which consists of two convolutional layers and a linear layer with residual connection, convert the text embeddings to hidden representation.
Meta-Voice adopts 4 feed-forward transformer (FFT) blocks for both the text encoder and mel decoder, following model design in FastSpeech2.
The variance adaptor has similar network architecture to FastSpeech2, where phoneme-level pitch and energy features are used.
A fully-connected (FC) layer is added on the output of FFT blocks in mel decoder to output raw predicted mel spectrogram. Inspired by \cite{shen2018natural}, we add a residual postnet to post-process raw mel spectrogram to obtain the final predicted mel spectrogram.
Sinusoidal positional encodings \cite{vaswani2017attention} is added to both the text encoder and the mel decoder following common practice. 

\subsection{Style-related parameters}
\label{sec:2-2}

The style encoder used in Meta-Voice is illustrated in Fig.~\ref{fig:style_encoder}.
Speaker information is represented by speaker IDs, while prosody information is extracted from mel spectrogram with a prosody mel encoder.
One-hot speaker IDs are transformed into speaker vector via a trainable speaker look-up-table (LUT), a speaker conditional layer normalization (CLN) adaptor consumes speaker vectors to generate speaker-related SALN parameters, i.e., scale $\gamma$'s and shift $\beta$'s.
The prosody mel encoder comprises of a spectral processing block, temporal processing block, a multi-head self-attention block and a final temporal averaging layer, following the design in \cite{min2021meta}, which convert a mel spectrogram into a fixed-length prosody vector. We then use a prosody CLN adaptor to generate prosody-related SALN parameters.

As shown in Fig.~\ref{fig:saln}, the SALN layer in Meta-Voice uses speaker-related SALN parameters $\{\gamma_s^i,\beta_s^i\}_{i=1}^M$ and prosody-related SALN parameters $\{\gamma_p^i,\beta_p^i\}_{i=1}^M$, where $M$ is the total number of SALN layers in Meta-Voice, to modulate hidden features in the text encoder and mel decoder, as
\begin{equation}
    \mathcal{H}_{out}^i = (\gamma_s^i \cdot LN(\mathcal{H}_{in}^i) + \beta_s^i) + (\gamma_p^i \cdot LN(\mathcal{H}_{in}^i) + \beta_p^i),
\end{equation}
where $LN(\cdot)$ denotes layer normalization without trainable affine transformation.

To fully disentangle information carried by speaker vector and prosody vector, we adopt two methods. We add a gradient reversal layer (GRL) and a speaker classifier on prosody vector, following the notion of domain adversarial training. We also add an orthogonal constraint on speaker vector and prosody vector following \cite{bian2019multi}. Both methods are used during the large-scale pre-training phase, while only the orthogonal constraint is used during 
the meta learning phase and the adaptation phase.

\section{Meta Learning Algorithm}
\label{sec3}

We adopt MAML as the meta learning algorithm.
The goal of MAML is to find a sensitive initial point of model parameters through cross-task meta-training such that a few gradient updates lead to large performance improvement in a new task during meta-test.
Previous works have shown that taking model parameters pre-trained on large-scale existing data as a warm start of meta-training can achieve better performance \cite{sun2019meta}, i.e., transferring knowledge from large-scale pre-training.
Therefore, in our case, the training process of Meta-Voice consists of three phases: large-scale pre-training, meta learning and adaptation to new speakers, as illustrated in Fig.~\ref{fig:overview}.

We denote a pair of text and mel spectrogram as $(X, Y)$.
During large-scale pre-training phase, parameters of Meta-Voice are initialized as $\theta_0=\{\bar{\theta}_0, \theta^{s}_0, \theta^p_0\}$ and optimized to minimize a reconstruction loss between a predicted mel spectrogram $\hat{Y}$ and the ground-truth $Y$, as well as the domain adversarial classification loss $\mathcal{L}_{da}$ and orthogonal loss $\mathcal{L}_{orth}$ introduced in Section~\ref{sec:2-2}:
\begin{equation}
\label{eq1}
\begin{split}
    \mathcal{L}^{\mathcal{D}}(\theta) &= \mathbb{E}[\|\hat{Y} - Y\|_1 + \alpha_1\mathcal{L}_{da} + \alpha_2\mathcal{L}_{orth}]
\end{split}
\end{equation}
where $\|\cdot\|_1$ is L1 distance and $\alpha_1$ and $\alpha_2$ are loss weights. After the pre-training phase, the initial model parameters $\theta_0$ becomes $\theta_T$.
We omit the reconstruction terms of pitch and energy features for brevity.

During the meta learning phase, we synthesize meta dataset $\mathcal{D}_{meta}$ for the cross-task training. The meta-learner is initialized with parameter $\theta_T$ and optimized with $\mathcal{D}_{meta}$, which is divided into $\mathcal{D}_{tr}$ for task-level training and $\mathcal{D}_{te}$ for task-level test.
We freeze shared parameters $\bar{\theta}$ and prosody-related parameters $\theta^p$ and only update speaker-related parameters $\theta^s$.
For an $N$-shot adaptation task $\mathcal{T}_i$, e.g., voice cloning with $N$ samples, we sample $N$ samples from a speaker to form $\mathcal{D}_{tr}$ and another $N$ samples from the same speaker to form $\mathcal{D}_{te}$.
In each inner loop step, a base-learner is initialized from the meta-learner, and conducts one or more gradient descent updates. For one gradient update, new adapted parameters $\theta_i$ is computed as:
\begin{equation}
    \theta^s_i=\theta^s-\alpha\nabla_{\theta^s}\mathcal{L}_{\mathcal{T}_i}^{\mathcal{D}_{tr}}(\theta),
\end{equation}
where $\alpha$ is the fast adaptation learning rate. During meta learning, the parameters of the meta-learner are trained to optimize the performance of the adapted model on the test split $\mathcal{D}_{te}$. 
Parameters of the meta-learner is updated by collecting a batch of task-level test losses as:
\begin{equation}
\theta^s \gets \theta^s - \beta\nabla_{\theta^s}\sum_{\mathcal{T}_i\sim p(\mathcal{T}_i)}\mathcal{L}_{\mathcal{T}_i}^{\mathcal{D}_{te}}(\{\bar{\theta}, \theta^{s}_i, \theta^p\})
\end{equation}
where $\beta$ is the meta learning rate.

The large-scale pre-training and meta learning procedures are summarized in Algorithm~\ref{alg:training}. After meta-transfer learning, we obtain model parameters $\theta_M$, which are expected to adapt to a new speaker using limited adaption data rapidly. 

\begin{algorithm}[h]
\caption{Large-scale Pre-training and Meta Learning}
\textbf{Input}: Large-scale multi-speaker and multi-prosody TTS corpus $\mathcal{D}$; Learning rate $\alpha$ and $\beta$ \\
\textbf{Output}: Meta model $\theta_M$
\begin{algorithmic}[1]
\label{alg:training}
\STATE Randomly initialize $\theta\gets\{\bar{\theta}_0, \theta^{s}_0, \theta^p_0\}$;
\FOR{\textit{samples} in $\mathcal{D}$}
\STATE Evaluate $\mathcal{L}^\mathcal{D}(\theta)$ by Eq.~\ref{eq1};
\STATE Optimize $\theta$ with gradient descent;
\ENDFOR
\STATE Generate task distribution $p(\mathcal{T})$ with $\mathcal{D}$
\WHILE{\textit{not done}}
\STATE Sample task batch $\mathcal{T}_i\sim p(\mathcal{T})$;
\FOR{\textit{all} $\mathcal{T}_i$}
\STATE Evaluate training loss $\mathcal{L}_{\mathcal{T}_i}^{\mathcal{D}_{tr}}(\theta)$;
\STATE Adapt speaker-related parameters with gradient descent: $\theta^s_i\gets\theta^s-\alpha\nabla_{\theta^s}\mathcal{L}_{\mathcal{T}_i}^{\mathcal{D}_{tr}}(\theta)$
\ENDFOR
\STATE Update $\theta^s$ with respect to average test loss: 
\STATE $\theta^s \gets \theta^s - \beta\nabla_{\theta^s}\sum_{\mathcal{T}_i\sim p(\mathcal{T}_i)}\mathcal{L}_{\mathcal{T}_i}^{\mathcal{D}_{te}}(\{\bar{\theta}, \theta^{s}_i, \theta^p\})$
\ENDWHILE
\end{algorithmic}
\end{algorithm}

\section{Experiments}
\label{sec4}

\subsection{Setups}

We use an internal multi-speaker and multi-prosody Mandarin Chinese corpus for both the pre-training and meta learning phases. The corpus contains data from 140 speakers, with 10-prosody annotations (i.e., \textit{neutral, happy, angry, sad, scary, news, story, broadcast, poetry and call-center}). In total, the corpus has 71 hours speech data.
During meta testing (i.e., adaptation phase), we use another speech dataset consisting of 4 neutral speakers (2 females and 2 males) to evaluate few-shot style transfer performance of Meta-Voice. Audio is sampled at 24 KHz.

Meta-Voice uses 4 feed-forward Transformer (FFT) blocks in both the text encoder and the mel decoder. The hidden size, number of attention heads, kernel size and filer size of the 1-dimensional convolution in the FFT block are set as 256, 2, 9 and 1024, respectively. The dimension of the speaker vector and the prosody vector is 128.

During the pre-training phase, we follow the common multi-speaker training procedure with loss function in Eq. (2), where we set $\alpha_1=0.01$ and $\alpha_2=0.02$.
In this work, we focus on 5-shot (around 12 second speech data in total) style transfer.
Therefore, during the meta learning phase, we sample 5 training samples and 5 test samples from one speaker to form a meta task. We regard a pair of speaker-prosody as a pseudo-speaker. Thus, in total there are 1400 pseudo-speakers, which are split into a meta training set (1300 pseudo-speakers) and a meta validation set (100 pseudo-speakers). We randomly initialize a speaker vector before meta learning, i.e., the speaker LUT only has one shared entry across all pseudo-speakers.
We use a meta batch size of 10. Meta learning rate is 0.0001 and base learning rate is 0.001. The Adam optimizer \cite{kingma2014adam} is adopted as the meta optimizer, while the vanilla SGD optimizer is adopted as the base optimizer.

A HifiGAN vocoder \cite{hifigan} is used to generate waveform from mel spectrogram, which is trained with the same multi-speaker and multi-style corpus.
During meta testing (i.e., adaptation process), the initial base learning rate is 0.001. We found that using an exponential learning rate scheduler with decay factor 0.9998 accelerates adaptation.

\subsection{Evaluations}

To the best of our knowledge, there is no existing work addressing few-shot style transfer in voice cloning.
We use the popular pre-train-finetune adaptation pipeline as the baseline. Specifically, the baseline model adopts the same model structure as Meta-Voice with the same pre-training procedure. During finetuning, we randomly initialize a speaker vector for a new speaker and update the speaker-related parameters only with the 5 adaptation samples. The adaptation process is the same as that of Meta-Voice.

In this section, we use the metrics of speaker cosine similarity and mel cepstral distortion (MCD) to objectively measure the few-shot voice cloning performance in terms of voice similarity and quality, respectively.
Audio samples are available online  \footnote{\url{https://liusongxiang.github.io/meta-voice-demo/}}.

We adopt a pre-trained speaker verification model to compute embedding vectors of the generated samples and their corresponding reference recordings and then compute paired cosine distances. Each compared model is evaluated with 5600 generated samples under all ten prosody annotations. MCDs are computed with only the neutral samples since the four testing speakers only have neutral recordings. 

\begin{figure}[t]
	\centering
	\includegraphics[width=0.5\textwidth]{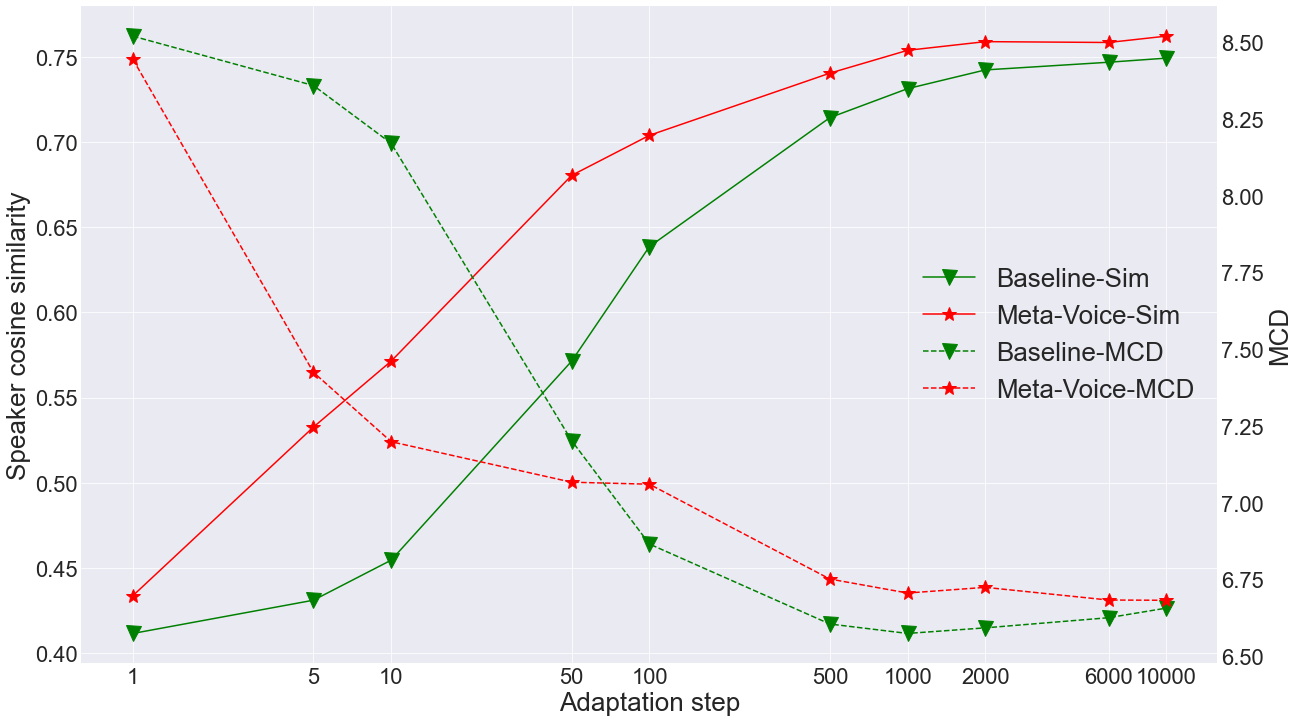}
	\caption{5-shot cross-gender style transfer results.}
	\label{fig:cg}
\end{figure}

\begin{figure}[t]
	\centering
	\includegraphics[width=0.5\textwidth]{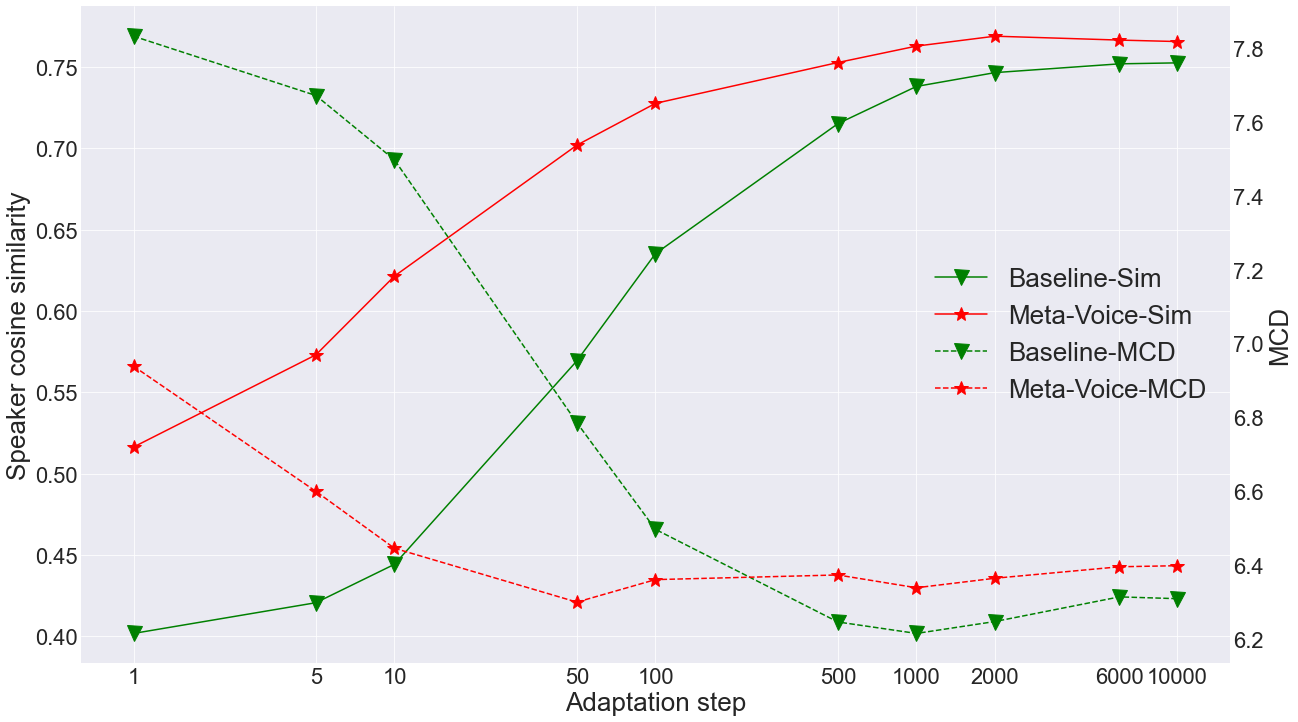}
	\caption{5-shot intra-gender style transfer results.}
	\label{fig:ig}
\end{figure}

5-shot cross-gender and intra-gender style transfer results are shown in Fig.~\ref{fig:cg} and Fig.~\ref{fig:ig}, respectively. We can see that Meta-Voice can adapt to a new speaker very quickly with speaker cosine similarity above 0.7 with only 100 adaptation steps, while the baseline needs much more adaptation steps to reach the same level of speaker cosine similarity. We see similar phenomena in MCD results.
However, it is worthy to note that the baseline model reaches similar level of speaker cosine similarities and MCDs to Meta-Voice after sufficient number of adaptation steps. Therefore, the major benefit of meta learning algorithms, such as MAML, is to learn a better model initialization such that the model can be quickly adapted to new speakers. Faster adaptation can significantly reduce the compuational cost and the overall processing time.

\section{Conclusion}
\label{sec5}

In this paper, we have presented a novel few-shot style transfer approach, named Meta-Voice, for voice cloning. The MAML algorithm is adopted to accelerate the adaptation process.
Domain adversarial training mechanism and orthogonal constraint are adopted to disentangle speaker and prosody representations for effective cross-speaker style transfer.
Experimental results have shown that Meta-Voice can achieve speaker cosine similarity of 0.7 within 100 adaptation steps for both cross-gender and intra-gender style transfer settings.
Future directions include investigating the use of meta learning methods in totally end-to-end TTS models (i.e., text-to-waveform synthesis).

\bibliographystyle{IEEEbib_abbrev}

\bibliography{strings,refs}

\end{document}